\documentclass[sigconf]{acmart}

\usepackage{booktabs} % For formal tables
\usepackage{balance}
\makeatletter
\def\@copyrightspace{\relax}
\makeatother

\renewcommand\footnotetextcopyrightpermission[1]{} % removes footnote with conference information in first column

\begin{document}
\title{FlyNetSim: An Open Source Synchronized UAV Network\\ Simulator based on ns-3 and Ardupilot}
\titlenote{This article has been accepted for publication in the  ACM MSWiM 2018. 21st ACM International Conference on Modelling, Analysis and Simulation of Wireless and Mobile Systems (MSWIM '18), October 28-November 2, 2018, Montreal, QC, Canada}
%\thanks{This article has been accepted for publication in the  ACM MSWiM 2018}
%\subtitlenote{This article has been accepted for publication in the  ACM MSWiM 2018}

\author{Sabur Baidya, Zoheb Shaikh  and Marco Levorato}
%\authornote{The secretary disavows any knowledge of this author's actions.}
\affiliation{%
  \institution{Donald Bren School of Information and Computer Sciences, UC Irvine, CA, USA}
  %\city{Irvine}
  %\state{CA}
  %\country{USA}
}
\email{[sbaidya, zsshaikh, levorato]@uci.edu}

\begin{abstract}
Unmanned Aerial Vehicle (UAV) systems are being increasingly used in a broad range of applications requiring extensive communications, either to interconnect the UAVs with each other or with ground resources. Focusing either on the modeling of UAV operations or communication and network dynamics, available simulation tools fail to capture the complex interdependencies between these two aspects of the problem.
The main contribution of this paper is a flexible and scalable open source simulator -- FlyNetSim -- bridging the two domains. The overall objective is to enable simulation and evaluation of UAV swarms operating within articulated multi-layered technological ecosystems, such as the Urban Internet of Things (IoT). To this aim, FlyNetSim interfaces two open source tools, ArduPilot and ns-3, creating individual data paths between the devices operating in the system using a publish and subscribe-based middleware. The capabilities of FlyNetSim are illustrated through several case-study scenarios including UAVs interconnecting with a multi-technology communication infrastructure and intra-swarm ad-hoc communications.
\end{abstract}


\maketitle

%%%%%%%%%%%%%%%%%%%%%%%%%%%%
%%%%%%  INTRODUCTION  %%%%%%
%%%%%%%%%%%%%%%%%%%%%%%%%%%%

\section{Introduction}

Unmanned Aerial Vehicles (UAV) are attracting a considerable attention from the research community. Traditional applications range from surveillance to precision agriculture, environmental monitoring and disaster management \cite{sona2016uav,erdelj2017help}. Recent use cases emphasize the role of communications and networking in the overall picture, either to enable intra-swarm coordination or to interconnect the UAVs to ground resources \cite{natalizio2018advances,wu2018joint}. A new line of contributions deeply integrates UAVs into the communication infrastructure to extend wireless coverage ~\cite{shaikh2018robust}.

Clearly, deploying real-world UAV systems poses several challenges, especially in urban or densely populated environments. Thus, simulation is an important option to reduce the overhead of testing new solutions and architectures. However, there has been a limited amount of effort in developing simulation tools capable to accurately model the fine-grain operations and characteristics of both UAV and networking/Internet of Things (IoT) systems. Mature and popular tools focus on either one of these aspects, often oversimplifying the other. Recent contributions such as AVENS~\cite{marconato2017avens} and CUSCUS~\cite{zema2017cuscus} make a first effort in this direction, interlacing UAV and networking simulation tools. This paper builds on these first steps to create a comprehensive simulator -- FlyNetSim -- capturing the intricate interdependencies between the communication and network environment and UAV operations, such as sensing and navigation, and inner state dynamics (\emph{e.g.}, battery state). We pose a particular emphasis on urban IoT systems, where the UAVs are immersed in a multi-scale technological ecosystem conglomerating several wireless access technologies and communication strategies -- \emph{e.g.}, Wi-Fi, Long Term Evolution (LTE) and Device-to-Device (D2D) -- as well as backhaul wired networks and computation resources such as edge and cloud servers \cite{bonomi2012fog}. 
Our main objectives are:

\vspace{2mm}
\noindent
$\bullet$ Accurately model UAV operations and dynamics using a software-in-the-loop approach, where the data structures and control pipeline of UAV software are fully preserved. 

\vspace{1mm}
\noindent
$\bullet$ Accurately model a multi-scale multi-technology IoT communication environment and its interactions with the UAVs. 

\vspace{1mm}
\noindent
$\bullet$ Establish a one-to-one correspondence between UAVs and wireless nodes in NS-3, where the UAVs implement a full network stack supporting multiple network interfaces.

\vspace{1mm}
\noindent
$\bullet$ Preserve individual data paths from and to UAV sensors and controllers.

\vspace{1mm}
\noindent
$\bullet$ Provide a graphical user interface to visualize the status of the system and automatically generate UAV/network scenarios.

\vspace{1mm}
\noindent
$\bullet$ Support emulation, where real-world UAVs can perform on-board simulations of a surrounding network environment while flying.

\vspace{1mm}
\noindent
$\bullet$ Support computation and real-time data processing in-the-loop, for instance to implement and test edge computing architectures.

\vspace{1mm}
To accomplish these objectives, we take as starting point two open source simulators -- Network Simulator (NS-3) and ArduPilot -- and build a fully open source simulation environment for academic research. NS-3~\cite{henderson2008network} is an extremely popular tool, with a wide community contributing to extend its capabilities. ArduPilot~\cite{team2016ardupilot} is a widely used software and hardware-in-the-loop simulator, capable of modeling a broad range of unmanned vehicles characteristics in terms of navigation, control and mission planning.

FlyNetSim includes a middleware layer to interconnect the two simulators, providing temporal synchronization between network and UAV operations, and a publish and subscribe based framework~\cite{eugster2003many} to create end-to-end data-paths across the simulators. The middleware architecture we propose is lightweight, and enables FlyNetSim to simulate a large number of UAVs and support a wide range of IoT infrastructures and applications. We illustrate the capabilities of FlyNetSim through several case-study scenarios:

\vspace{2mm}
\noindent
$\bullet$ \emph{UAV control over Wi-Fi}: We use FlyNetSim to evaluate the effect of mobility, delay, and congestion on the operations of a UAV remotely controlled by a Ground Control Station (GCS).

\vspace{1mm}
\noindent
$\bullet$ \emph{Multi-network communications}: We create and test a case-study scenario in which a UAV uses multiple network technologies for uplink and downlink communications. 

\vspace{1mm}
\noindent
$\bullet$ \emph{D2D Communications for UAV swarms}: We evaluate through FlyNetSim the ability of D2D communications to extend network coverage through intra-swarm communications.

\vspace{1mm}
\noindent
$\bullet$ \emph{IoT and Data Streaming}: FlyNetSim supports real-time data streaming from the UAVs. We showcase this ability by testing real-time telemetry and video streaming.

Finally, we also showcase the ability of FlyNetSim to support the ``emulation mode'', where the simulator runs on a real-world UAV. 

\vspace{1mm}

The rest of the paper is organized as follows. Section~\ref{sec:related} reviews state-of-the-art simulation tools. In Section~\ref{sec:framework} and~\ref{sec:architecture}, we discuss challenges and design criteria and present the architecture of FlyNetSim. 
In Section~\ref{sec:eval} we describe the use case scenarios used to illustrate the capabilities of the proposed simulator and provide extensive numerical evaluations. Section~\ref{sec:emu} describes the on-board emulation capabilities of FlyNetSim. Section~\ref{sec:conclusions} concludes the paper.

%%%%%%%%%%%%%%%%%%%%%%%%%%%%%%%%
%%%%%%  State of the Art  %%%%%%
%%%%%%%%%%%%%%%%%%%%%%%%%%%%%%%%
\section{State of the Art}
\label{sec:related}

In this section, we provide an overview of UAV and network simulators, and discuss the recently proposed AVENS and CUSCUS.

\vspace{2mm}
\noindent{\bf UAV Simulators - }
The primary scope of UAV simulators is to model aerodynamics and functional aspects of UAV control systems. Early simulators, \emph{e.g.},~\cite{hsimuav}, only simulated human-controlled flight, and, autonomous flight and autopilot features were implemented in later tools. Among open source autopilot simulators, Ardupilot and PX4 are most widely used, with the former having a wider range of supported platforms and hardware, including Pixhawk~\cite{meier2012pixhawk}, NAVIO~\cite{naviosite}, and Erle-Brain~\cite{erlebrain}. Ardupilot is also used by some commercial vehicles, such as 3DR Solo~\cite{3drsolo}.

Several UAV simulators build on top of Ardupilot to extend its capabilities.  For instance, Gazebo~\cite{koenig2004design} extends the range of simulated devices to include land rovers, planes, and small robots using the Robot Operating System (ROS)~\cite{quigley2009ros}. However, Gazebo is a very heavy-weight software, which may be difficult to run on a single machine when simulating a large number of vehicles. Moreover,  ROS is not synchronized with real-time operations and communications of connected applications. XPlane-10, RealFlight and few other commercial simulators are also based on Ardupilot, but mostly focus on motion and navigation. To develop FlyNetSim, we choose Software In The Loop (SITL)~\cite{team2016sitl,ardusitl}, one of the simplest simulators based on Ardupilot supporting autopilot navigation.

\vspace{2mm}
\noindent{\bf Network Simulators - }
Many simulators are available which offer simulation of the full network stack or a portion of it. Most simulators are discrete event-driven, meaning that they form a timeline of events that is sequentially processed, without a strict correspondence between real and simulated time.
One of the earliest discrete event-driven simulator is TOSSIM~\cite{levis2003tossim} which was used to model communications between wireless sensors. Given our objectives, we are primarily interested in network simulators that can support WiFi and 4G/5G cellular communications. Also, the simulator should support a wide range of protocols and options at different layers of the network.

OPNET~\cite{chang1999network} is a commercial simulator providing detailed simulation of wireless networks and supports a wide spectrum of protocols and technologies. Among open source  network simulators OMNET++~\cite{varga2008overview} and NS-3~\cite{riley2010ns} are the most used by the research community. Both simulators are discrete-event driven. Thanks to a wide support from the community, NS-3 can simulate many technologies and embeds a variety of statistical models for channel gain, mobility and traffic generation, and provides a detailed modeling of physical layer operations. Additionally, NS-3 can efficiently interface with external systems, applications and libraries. For the reasons above, we choose NS-3 as the network component of FlyNetSim.

\vspace{2mm}
\noindent{\bf Joint UAV/Network Simulators - }
Recent work~\cite{marconato2017avens} integrates OMNET++ and X-Plane to build a joint UAV-Network simulator -- the Flying Ad Hoc Network Simulator (AVENS). The main limitation of AVENS is that the UAV and network simulator communicate via a XML file, which only updates the number and position of the UAVs. This communication model does not support data-paths from and to the UAVs to transfer control, telemetry and sensor data. Also, the UAV and network simulators are not synchronized, thus impairing the ability of AVENS to model and evaluate real-time services and operations. Finally, X-Plane is not an open-source software, thus making the use of AVENS challenging in academic research.

The joint UAV-network simulator CUSCUS~\cite{zema2017cuscus} has been developed based on FL-AIR~\cite{flair} in combination with NS-3 using tap bridges~\cite{alvarez2010limitations} and containers. CUSCUS establishes NS-3 communication links through Linux containers using real-time scheduling, thus successfully creating network devices corresponding to the UAVs in FL-AIR.
However, the current NS-3 implementation has limited support to interfaces through tap bridges. For instance, this approach prevents the inclusion of important technologies such as LTE and D2D. Also, tunneling through the tap interfaces does not facilitate multi-hop routing through intermediate nodes in the network simulator. These aspects impair the ability of CUSCUS to simulate important multi-technology environments and systems such as the urban IoT. 
Additionally, in the CUSCUS architecture the UAV control loop with the Ground Control Station (GCS) is discorporated from the network simulation. So, the simulator cannot be used to evaluate the effect of the network environment dynamics on the operations of the UAVs.
Finally, the use of containers to run the UAVs imposes a larger computational complexity compared to other options. This aspect is particularly relevant, because if NS-3 event processing is delayed compared to the UAV simulator clock due to a large number of network events, then synchronization will break. Therefore, an approach based on containers severely limits the complexity of the simulated environment.

\begin{figure}[!t]
\centering
\includegraphics[width=\columnwidth, height=0.65\columnwidth]{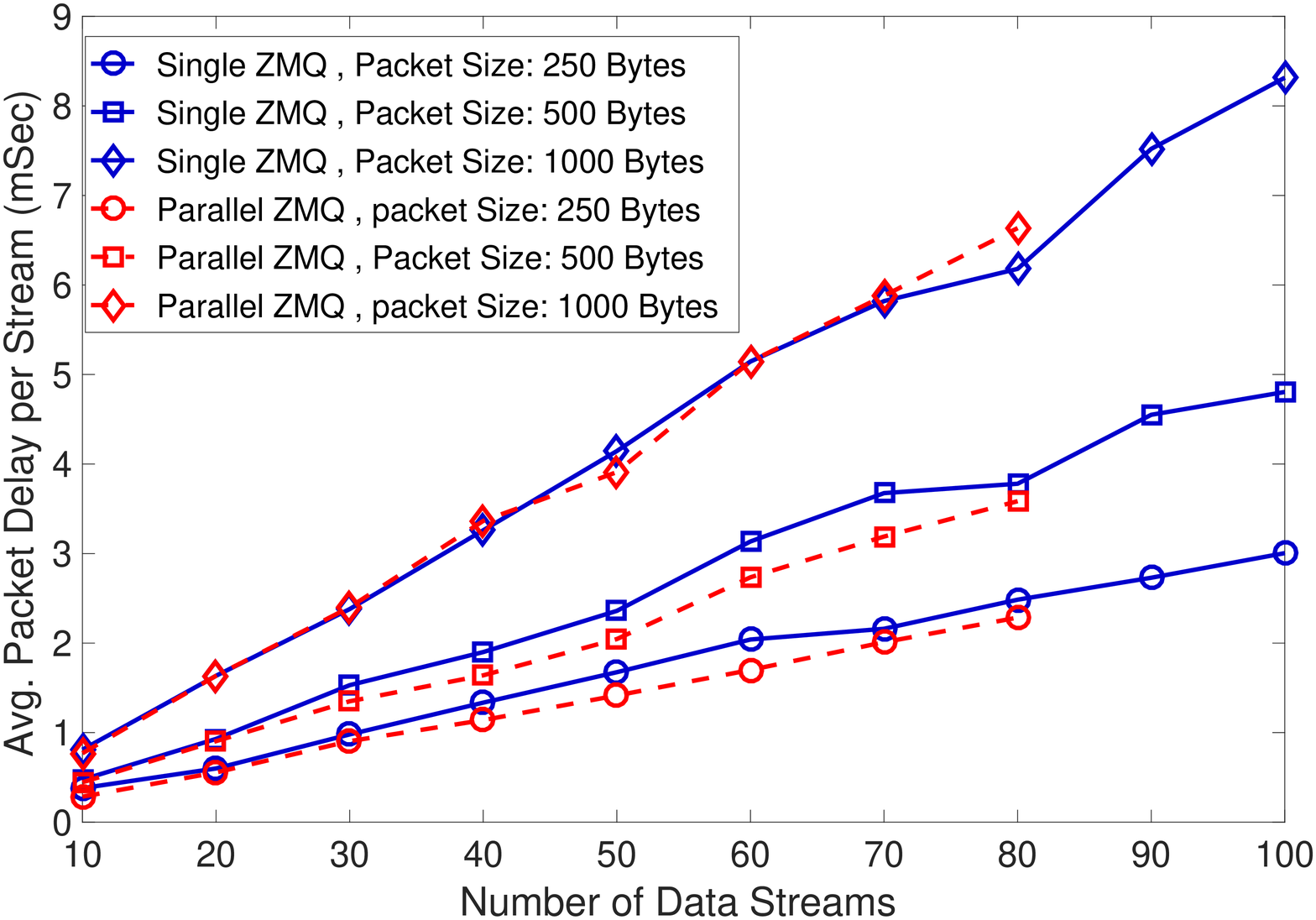}
\vspace{-4mm}
\caption{Average total ZMQ delay as a function of the number of nodes using a single or multiple ZMQ publisher/subscriber pairs.}
\label{fig:zmqs}
\vspace{-2mm}
\end{figure}

%%%%%%%%%%%%%%%%%%%%%%%%%%%%%%%
%%%%%% General Framework %%%%%%
%%%%%%%%%%%%%%%%%%%%%%%%%%%%%%%
\section{General Framework}
\label{sec:framework}

As indicated in the previous section, we choose NS-3 and ArduPilot as the two main blocks of FlyNetSym. However, interconnecting the two simulators is non-trivial. One of the key desiderata is to build a lightweight simulator capable to include a large number of UAVs and other devices, and a complex communication infrastructure. In the following, we discuss some of the available options and the chosen approach.

\vspace{2mm}
\noindent
{\bf Virtual Machines -} A possible approach is to
run each UAV in an individual Virtual Machine (VM) and connect the VMs via the network simulator running on the native host. However, this approach results in a considerable resource usage, which impairs scalability. Moreover, communication through the upper layer of the protocol stack of the host to the VMs introduces some delay and necessitates additional memory usage. Finally, due to the isolation of the environment embedding each UAV, their synchronization would require the implementation of a complex algorithm.

\vspace{2mm}
\noindent
{\bf Containers -}
The second option is to use lighter-weight containers instead of VMs, thus mitigating the scalability issue. Note that containers are used in  CUSCUS. However, containers share the synchronization issue with the VMs.  
Moreover, the containers would need to be connected to NS-3 via tap bridges (see the approach in CUSCUS ~\cite{zema2017cuscus}) associated with a ``ghost node" in NS-3. However, in the current implementation of NS-3, this communication model does not support several key technologies, thus limiting the range of possible simulation environments. Additionally, the use of containers would prevent ``on-board'' simulations, as the autopilot hardware would need to run within a container.

\vspace{2mm}
\noindent
{\bf Multi-threading -}
The option we select for the development of FlyNetSim is multi-threading, where each UAV simulator module corresponds to a thread. This model provides improved scalability and seamless synchronization between the UAVs and the network simulator. Additionally, on-board simulation will be possible, as the network simulator and the UAV run on the same host.

We remark that in this model we can directly create a unique node in NS-3 corresponding to each UAV and the GCS, without the need to have tap interfaces. Also, each UAV (NS-3 node) can be equipped with multiple network interfaces for multi-technology communication. 

The main challenge we face is the establishment of an effective communication between each UAV simulation module. A communication model based on files or shared memory makes the construction of individual and end-to-end data paths difficult due to the need for sophisticated mapping. This especially applies to a case where heavy-weight data streams need to be transported -- such as video or data streams -- a case in which file or memory sharing approach would grant inferior efficiency and performance. We  overcome this issue by incorporating Zero Message Queue (ZMQ)  in the framework -- a fully open source library based on a publish and subscribe communication model.

\vspace{1.5mm}
\noindent
\emph{\underline{ZMQ-based communication model}:}
ZMQ is a light weight, extremely fast messaging library written in C/C++ with binding support in various languages. This flexibility makes it a widely accepted tool to support cross-platform IoT applications. The sockets in ZMQ are created by the publishers and subscribers of the messages in the message queue of a broker. Thus, the publisher to subscriber connections are non-blocking and fully asynchronous. ZMQ also supports message filtering at the subscriber. 

Importantly, ZMQ supports many-to-many communication between end-points. This allows the establishment of sensor- and control-specific streams of traffic between the UAVs. However, it may result into an increased delay due to interleaving of packets in the queue, which in turn would result in an increased total end-to-end delay of the ZMQ communication. In order to assess this aspect, we perform an evaluation of the end-to-end delay of ZMQ defined as: 
\begin{equation}
D_{ze2e} = D_{pub} + D_{q} + D_{sub}
\end{equation}
where, $D_{ze2e}$ is the delay between a ZMQ sender and receiver, $D_{pub}$ is the publishing (sender to the ZMQ), $D_{sub}$ is the reception delay (ZMQ to the receiver) and $D_{q}$ is the queueing delay in the ZMQ.

\begin{figure*}[!t]
\centering
\includegraphics[width=1.9\columnwidth, height=0.9\columnwidth]{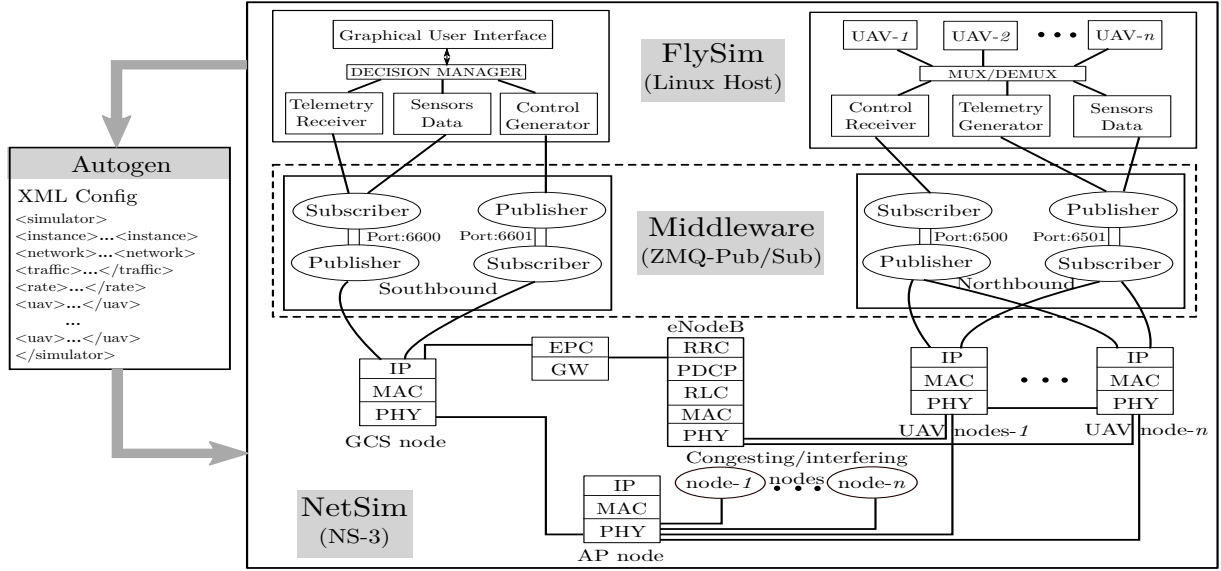}
\vspace{-1mm}
\caption{Architecture of FlyNetSim.}
\label{fig:uiot}
\vspace{-2mm}
\end{figure*}

We measure the total delay in two scenarios: \emph{(Single P/S)} the streams are sent over a single ZMQ publisher and subscriber pair or \emph{(Parallel P/S)} each data stream is matched with a ZMQ publisher and subscriber pair. Fig.~\ref{fig:zmqs} shows the end-to-end delay as a function of number of data streams $N$. As expected, the total delay increases with $N$. The single and parallel strategy have comparable delay, and in both cases the end-to-end packet delay is less than $10$ milliseconds even for large values of $N$ ($100$ streams) and data packet size ($1000$ bytes). Thus, both options have sufficiently small delay to support the simulation environment. However, a disadvantage of the parallel strategy is that each data stream has an individual socket pair, and a corresponding file descriptors for each of ZMQ end. In our preliminary study, we could create up to 80 parallel ZMQ publisher/subscriber pairs. Note that there is default limit of a single machine that restricts the number of file descriptors. The default value can be increased at the cost of an increased memory usage. Herein, we use a single ZMQ publisher/subscriber pair per communication direction for multiple data streams. This choice can support hundreds of concurrent data streams with lower complexity in implementation.

%%%%%%%%%%%%%%%%%%%%%%%%%%%%
%%%%%%  ARCHITECTURE  %%%%%%
%%%%%%%%%%%%%%%%%%%%%%%%%%%%
\section{Architecture}
\label{sec:architecture}

The architecture of FlyNetSim as depicted in Fig.~\ref{fig:uiot}, consists of the following major components: the \emph{FlySim} component that initiates the UAVs, GCS, and the visualization tools; the \emph{NetSim} component simulates the end-to-end network infrastructure and environment (including non-UAV nodes); the \emph{Autogen} component creates a map between the UAV components (including sensors) and the corresponding network nodes, and the interfaces and other infrastructure needed by the NetSim; and a \emph{Middleware} component supporting ZMQ-based bidirectional communications through the network. In the following, we describe in detail these components.

\vspace{2mm}
\noindent
\underline {\bf{\emph{1) FlySim:}}}
FlySim is the top layer of the architecture, and  contains the UAV and GCS node(s) running on the native host machine. The module instantiates the number of UAVs defined in the user input. Each instantiated UAV corresponds to a simulated vehicle in the SITL simulator based on Ardupilot. Thus, each vehicle is accompanied by a vector state indicating variables such as position, battery state, telemetry and sensor information. The GCS side contains a custom graphical user interface (GUI) to configure and control the UAVs to be created, and track their positions and other information from the telemetry. 

\vspace{2mm}
\noindent
\underline {\bf{\emph{2) NetSim:}}}
The NetSim component creates an interface with NS-3, integrated with ZMQ and other necessary libraries based on the configuration of communications in the simulated environment. The interface creates nodes such that there is one-to-one correspondence between each UAV in Flysim and a node in NetSim. Moreover, it enables the UAVs to be equipped with multiple wireless interfaces, \emph{e.g.}, WiFi and LTE, D2D. NetSim also configures the network environment in terms of congestion and interference from nodes outside of FlySim based on user input. The user can define the number of interfering nodes, their position, traffic rate, data statistics and technology. The module allows the use of any wireless protocol or strategy supported by NS-3, including ad-hoc point-to-point communications between the UAVs or infrastructure based communications over technologies such as LTE or WiFi. Finally, the NetSim module also handles mobility within the wireless network based on the actual mobility of the 
UAVs retrieved from the FlySim.

\vspace{2mm}
\noindent
\underline {\bf{\emph{3) Autogen:}}}
The Autogen component is responsible for automatic code generation in the FlySim and NetSim based on the input defined by the user in the graphic interface within the GCS. The module creates an XML file based on the input, which is passed to the NetSim as its arguments to create the nodes. 

\vspace{2mm}
\noindent
\underline {\bf{\emph{4) Middleware:}}}
The middleware interconnects the FlySim and NetSim via communication over ZMQ, creating the end-to-end data path for individual data streams from the GCS to the UAVs and vice versa. As both the GCS and the UAV application softwares are located outside the NetSim, the middleware needs to be used at both ends of the NetSim. We name the parts of the middleware as north-bound and south-bound middleware. The north-bound part connects the NetSim to the UAVs, whereas the south-bound connects the GCS to the NetSim. As described in the previous section, we use one ZMQ publisher/subscriber pair for communication in each direction for each of north-bound and south-bound middleware. The middleware is also responsible for synchronization between the FlySim and NetSim block as described in the following.

\subsection{Synchronization}

NS-3 is a discrete event-driven simulator, whose scheduler typically executes the events sequentially without synchronizing with an external clock. On the other hand, ArduPilot is a real-time operator, where the temporal evolution of the UAVs' state follows the external clock. To solve this disparity, we use the real-time scheduler in NS-3, which aligns the processing of scheduled events with their actual time. 
However, we remark that in dense simulations the scheduler may actually be ``late'' with respect to the clock. In this case, either synchronization is lost (``Besteffort") or late events are discarded (``Hardlimit").

\begin{figure}[!t]
\centering
\includegraphics[width=\columnwidth]{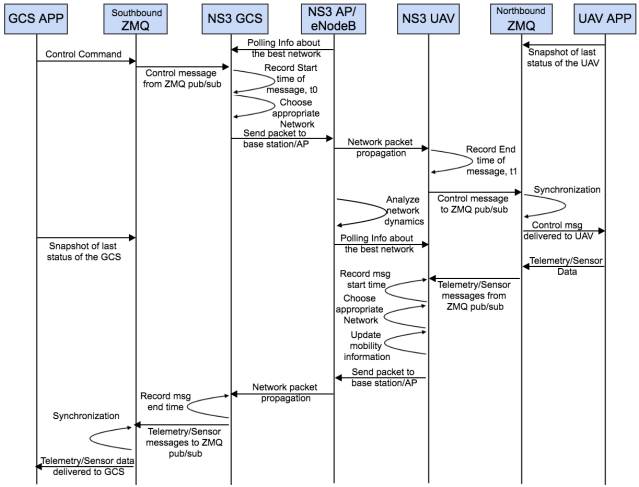}
\vspace{-2mm}
\caption{Message flow across the FlyNetSim architecture}
\label{fig:mflow}
\vspace{-3mm}
\end{figure}

We embed in FlyNetSim middleware's additional steps to ensure synchronization between the two simulators by timestamping each packets flowing through NS-3.  Fig.~\ref{fig:mflow} shows the message flow across the different layers of the simulators to preserve synchronization between the network and UAV operations. 
Let $t_{0}$ be state of the system clock at the time when a packet enters the NS-3 environment, and $\Delta$ the difference between the submission and reception time, that is, the network propagation time. When the packet exits NS-3, we acquire the system time $t_{1}$. If $(t_{1} - t_{0})$ is smaller than $\Delta$, \emph{i.e.}, the network event processing is faster than real-time, the middleware waits an additional $(\Delta + t_{0} - t_{1} )$ interval to release the packet to ArduPilot. 

In the case in which the NS-3 scheduler falls behind real-time, we propose to ``\emph{freeze}'' the UAV simulator to match the elapsed time with the current clock time. Note that this is critical to preserve an accurate reconstruction of flight dynamics and inner state (\emph{e.g.}, residual battery charge) evolution with respect to network events.
The idea is to take a snapshot of the vehicle state after the completion of last command and recreate the simulation at the correct time. This will avoid the simulator performing actions for a longer time than prescribed.

\subsection{Mobility}

FlyNetSim matches the actual dynamics of the UAV in the FlySim with the position of the corresponding wireless nodes in NS-3.
This is realized by intercepting the variables indicating the position, speed and direction in the telemetry flow at its creation, thus avoiding the, possibly large, time needed to receive the actual telemetry packets through the network. Note that telemetry is not created in response to commands, but, instead, in response to any update in the state of the vehicles. This ensures an accurate matching between the two simulators, where the network counterpart has the same granularity of representation of UAV dynamics as the UAV simulator.
Note that non-UAV nodes in NS-3 can use available mobility models, such as constant position, constant speed, random walk, \emph{etc.}

%%%%%%%%%%%%%%%%%%%%%%%%%%%%
%%%%%% IMPLEMENTATION  %%%%%
%%%%%%%%%%%%%%%%%%%%%%%%%%%%
\subsection{Implementation}
\label{sec:impl}

The implementation of FlyNetSim is based on C/C++ and Python. Specifically, we integrate the  ``libzmq" Python library with the UAV simulator and the ``libczmq" library bindings with NS-3. The network simulator is realized with two parallel threads which continuously listen for messages from the GCS and UAVs, respectively. The module within the NS-3 code that sets the position of the UAVs based on the telemetry data received from the external ZMQ in the listening thread, is based on the Haversine formula~\cite{veness2011calculate}.

On the application side, each UAV is instantiated using a thread running the UAV simulator based on ArduPilot. We use Dronekit~\cite{dronekit2015} to operate the UAVs following the Micro Air Vehicle Link (MAVLink) protocol~\cite{meier2013mavlink} that includes pitch, roll, and yaw movements. The custom GCS is created along with a GUI written in Python that can be easily integrated with ``pymavlink" libraries as well as receiving telemetry informations using MAVLink.

%%%%%%%%%%%%%%%%%%%%%%%%%%%%%%%%%%%%%%%%%%%%%%%%%%%%%%%
%%%%%%%% Case Studies and Numerical Evaluation  %%%%%%%
%%%%%%%%%%%%%%%%%%%%%%%%%%%%%%%%%%%%%%%%%%%%%%%%%%%%%%%
\section{Case Studies and Numerical Evaluation}
\label{sec:eval}

We provide a series of case-study scenarios to illustrate the capabilities of FlyNetSim to capture fine-grain interdependencies between communication/networking and UAV operations, as well as to simulate a wide range of rich network and IoT environments.

\subsection{Case Study I: Single UAV over WiFi}

In this first case study scenario, we consider a single UAV communicating with the GCS over WiFi. This simple scenario has been modeled using other integrated UAV/network simulators to evaluate the effect of mobility on bidirectional communications. However, FlyNetSim allows the inclusion of other nodes injecting traffic into the network, thus creating congestion.

\begin{figure}[!t]
\centering
\includegraphics[width=\columnwidth, height=0.8\columnwidth]{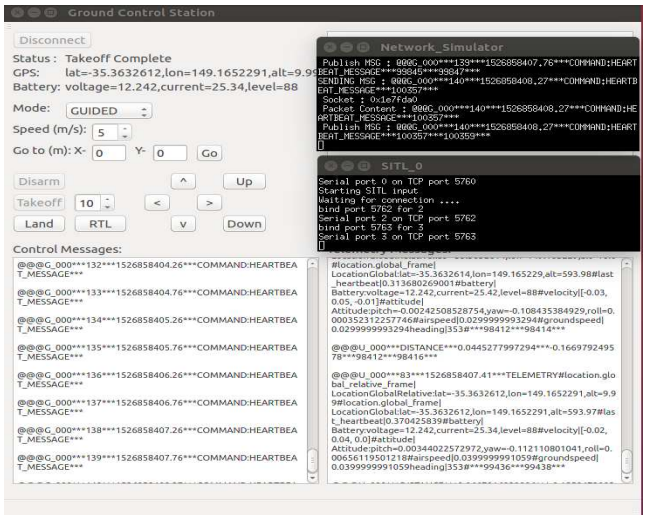}
\vspace{-4mm}
\caption{GUI Panel for Controls and Telemetry logs}
\label{fig:uavgui}
\vspace{-2mm}
\end{figure}

First, we assume the UAV hovers to maintain a constant position at a fixed distance from the WiFi Access Point (AP). The simulation accurately models the control messages sent from the GCS to a single instance of UAV through the GUI panel as depicted in Fig.~\ref{fig:uavgui}. The figure also shows the control and telemetry logs on the GUI. In the absence of other traffic, the average packet propagation delay through the network is approximately $0.24$~ms.

\begin{figure}[!t]
\centering
\includegraphics[width=\columnwidth, height=0.6\columnwidth]{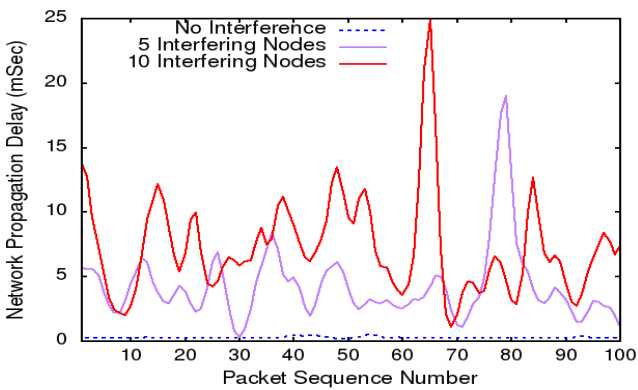}
\caption{Temporal trace of network delay in response to variations in channel quality and availability due to contention.}
\label{fig:delsmooth}
\vspace{-1mm}
\end{figure}

\begin{figure}[!t]
\centering
\includegraphics[width=\columnwidth]{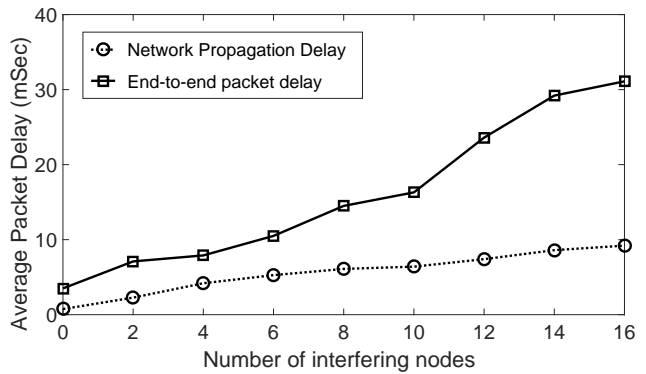}
\caption{Network and end-to-end delay as a function of the number of nodes contending for the channel (excluding the UAV).}
\label{fig:delc}
\vspace{-1mm}
\end{figure}

We measure the delay in the presence of nodes contending for the channel resource in the WiFi network, where each node has a traffic arrival rate of $10$~Mbps and each packet is $1000$~bytes long. 
Fig.~\ref{fig:delsmooth} shows a temporal trace of the network delay of control packets generated at the GCS and propagating through the network to the UAV. We use adaptive data rate for the WiFi channel in NS-3. 
In the absence of interference, the total delay is below $1$~ms. When the number of contending nodes is set to $5$ and $10$
the delay increases to an average of approximately $4.20$~ms and $7.06$~ms, respectively. Note that the delay presents significant variations over time due to channel contention and adaptive data rate, where the delay variance increases with the number of contending nodes. This effect may create undesirable jitter in the reception of control packets. 
Fig.~\ref{fig:delc} depicts the average delay of control packets reception as a function of the number of nodes contending for channel access.

\begin{figure}[!t]
\centering
\includegraphics[width=\columnwidth]{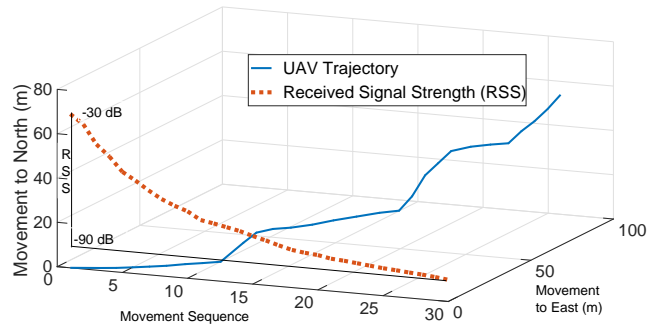}
\caption{Variation of received signal strength (WiFi) in response to the motion of the UAV.}
\label{fig:mobrss}
\vspace{-1mm}
\end{figure}

In the same setup, we test the feature of FlyNetSim matching the position of the UAV in NS-3 to that in the UAV simulator. For this experiment, we use random movements of the UAV from the GUI panel, and receive updated latitude and longitude values from the telemetry. The NetSim component of FlyNetSim converts the position in distance from the other nodes. We measure the received signal strength (RSS) of the UAV node from the AP. The blue line in the Fig.~\ref{fig:mobrss} shows the trajectory of the UAV over time in the XY plane and the dotted red line shows the corresponding decrease in the RSS value. The fine-granularity correspondence of the RSS in the simulator to the position of the UAV determined by its operations enables the development and testing of  physical and link layer protocols.


\subsection{Case Study II: Multi-Network Environment}

FlyNetSim seamlessly supports heterogeneous wireless network environments, as well as multiple wireless interfaces at each UAV. To the best of our knowledge, FlyNetSim is the only integrated simulator including this feature. 

To illustrate this capability, we implemented a multi-network environment including WiFi and LTE networks, where each UAV has a WiFi and an LTE interface and the corresponding protocol stack. In this use case, the GCS can be placed at the network edge, \emph{i.e.}, connected to the WiFi AP and the LTE eNodeB. Herein, we characterize the GCS as a remote host connected to the Evolved Packet Core (EPC) gateway of the LTE network to support long-range communication. The autogen module of the simulator creates the multiple wireless interfaces as defined by the input.

We focus on a scenario where the WiFi network is used to transport control messages from the GCS to the UAVs, and the LTE network is used to transport telemetry and high volume sensor data traffic from the UAV to the GCS. Note that case-studies where the network is dynamically selected based on Quality of Service (QoS) requirements can be implemented.

\begin{figure}[!t]
\centering
\includegraphics[width=\columnwidth, height=0.6\columnwidth]{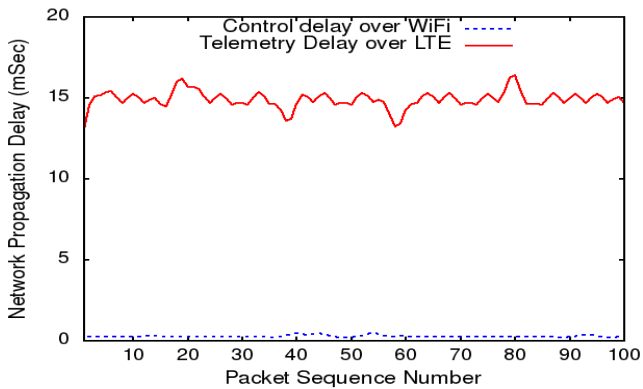}
\caption{Network delay of control messages (WiFi) and telemetry (LTE).}
\label{fig:lat_dual}
\vspace{-1mm}
\end{figure}

In the shown experiments, the WiFi nodes use adaptive data rate with maximum rate of $54$~Mbps. The LTE network is created using different uplink and downlink frequency bands of $20$~MHz each, and the LTE Frequency Division Duplex (LTE-FDD) mode is used for transmission. Nakagami channel model is used to model LTE signal propagation~\cite{yip2000simulation}. Specifically, we adopt the trace-based Extended Pedestrian A (EPA) fading model as indicated in the 3GPP standard~\cite{lte2009evolved}. The LTE network determines the Channel Quality Index (CQI) and assigns the corresponding Modulation and Coding Scheme (MCS) for transmission. Fig.~\ref{fig:lat_dual} depicts the end-to-end delay of the control data (transmitted over WiFi) and telemetry data (transmitted over LTE), which average at approximately at $0.24$~ms (control packets) and $15$~ms (telemetry). We remark that telemetry information is extracted before its transmission and immediately used to update the position of the UAV, which determines the channel gain of both LTE and WiFi channels.

\subsection{Case Study III: Multi-UAV}

This case-study scenario focuses on the simulation of UAV swarms. First, we compare the scalability of the proposed approach with that of the approach adopted by AVENS and CUSCUS. Note that the number of UAVs in AVENS is limited to $20$ due to restrictions in X-Plane, whereas CUSCUS does not have any strict limit, and was used to simulate up to $30$ UAVs.

We tested the scalability of FlyNetSim by instantiating multiple UAVs with basic functionalities, including the transmission of control messages from the GCS to each UAV and telemetry on the reverse path. We run the simulator on a laptop with $16$~GB of RAM and an Intel Core-i7-6700HQ processor. We could instantiate up to $70$ UAVs communicating using a WiFi network. The maximum number of UAVs is $40$ when LTE network is used, due to limitations imposed in NS-3 in connection with the $40$~ms periodicity of transmissions. However, the value can be increased up to $380$ based on the standard. Our architecture does not impose any restriction on the number of UAVs, and is only constrained by the available system resources. Therefore, more than $70$ UAVs can be instantiated with a more powerful computer. 

\begin{figure}[!t]
\centering
\includegraphics[width=\columnwidth]{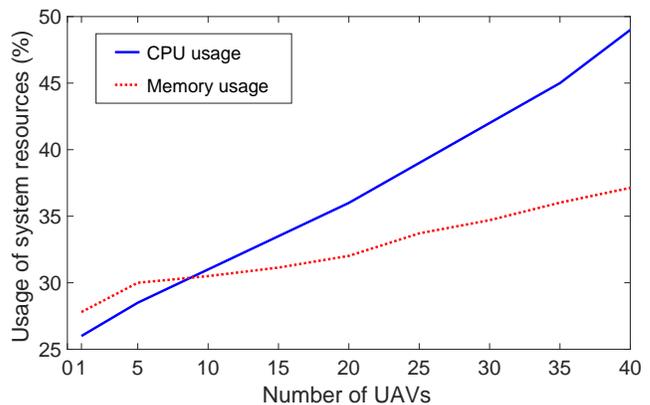}
\caption{Percentage of system resource usage as a function of the number of UAVs instantiated in the simulator.}
\label{fig:res_util}
\vspace{-1mm}
\end{figure}

We measure the system resource consumption in terms of memory and CPU usage as a function of number of UAVs instantiated. Fig.\ref{fig:res_util} shows that both the CPU usage and memory usage increase linearly with the number of UAVs. A direct comparison with AVENS simulator is not possible. However, we compare FlyNetSim resource usage to that of CUSCUS, which is built on network containers: FlyNetSim almost halves CPU usage and substantially reduces memory usage.

\begin{figure}[!t]
\centering
\includegraphics[width=\columnwidth, height=0.6\columnwidth]{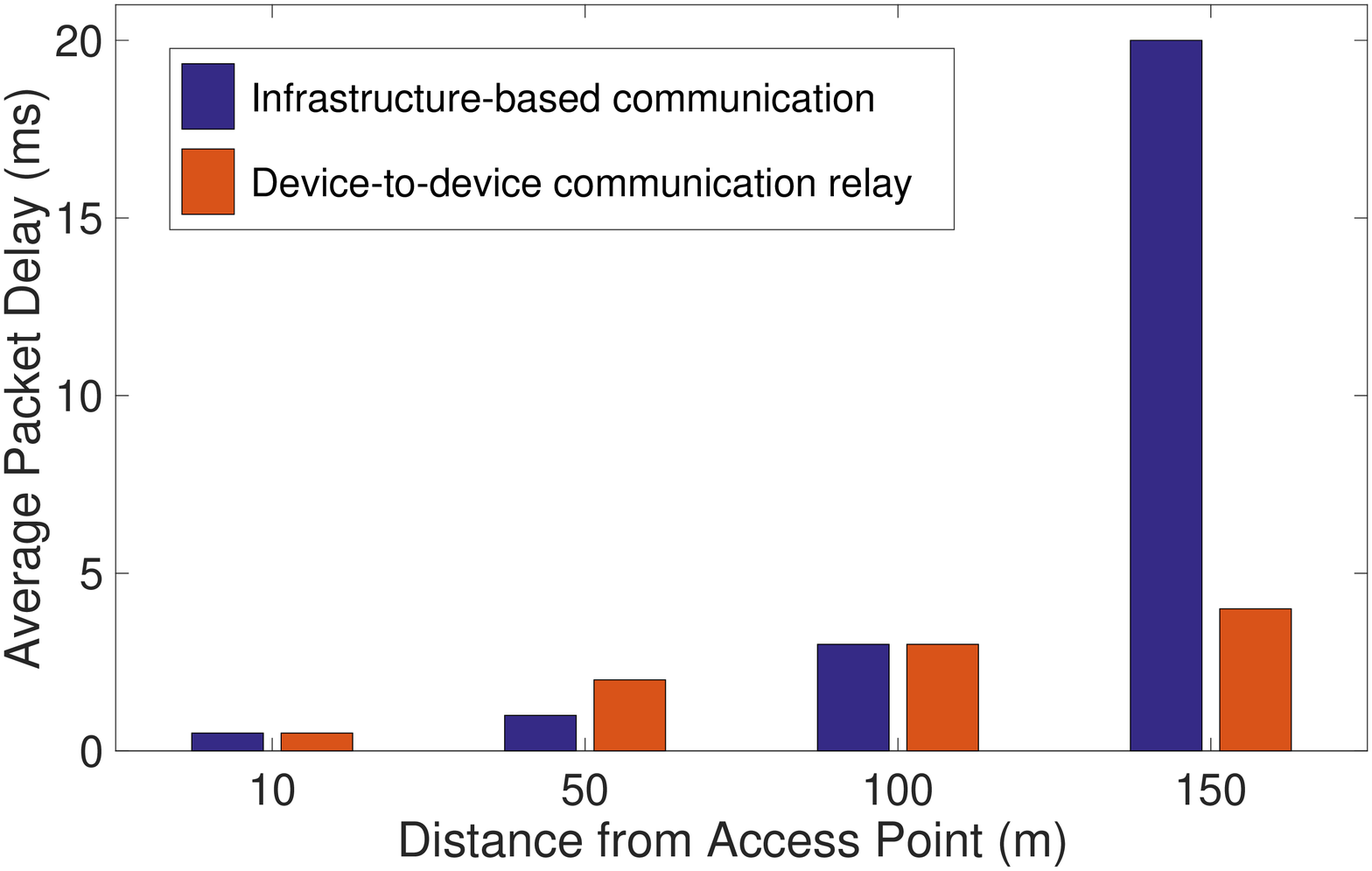}
\caption{Network delay as a function of the distance between the closest UAV in the line topology and the WiFi access point.}
\label{fig:comrelay}
\vspace{-1mm}
\end{figure}

We further demonstrate the capabilities of FlyNetSim by instantiating a use-case where the UAV swarm collaboratively transports packets to extend network coverage. The UAVs use D2D communications to form an ad-hoc network and relay messages from and to the GCS.
In the experiment we created $4$ UAVs positioned in a line topology where the UAVs are $50$~m apart from each other. We then measure the network delay as a function of the distance between the WiFi AP and the closest UAV. Fig.~\ref{fig:comrelay} shows that if packets are directly transmitted to the UAVs the average packet delay considerably increases as the UAVs approach the border of network coverage. Intra-swarm D2D
communications considerably extend coverage. Note that a similar scenario was shown in CUSCUS, where the UAVs broadcast the messages to the swarm, instead of forming an ad hoc network based on D2D communications.

\begin{figure}[!t]
\centering
\includegraphics[width=\columnwidth]{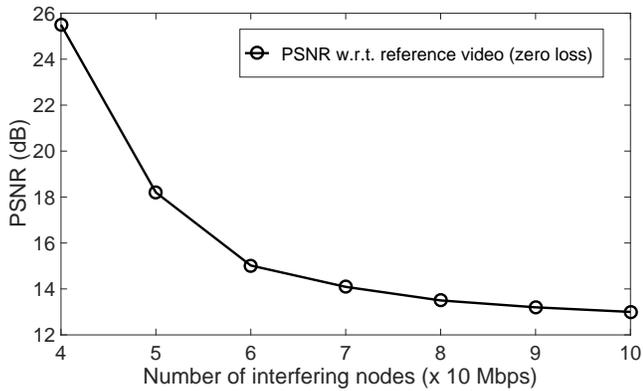}
\caption{PSNR as a function of the number of nodes contending the channel in the WiFi network.}
\label{fig:psnr_vid}
\vspace{-1mm}
\end{figure}

\subsection{Case Study IV: IoT Applications}

Unlike existing integrated network/UAV simulators, FlyNetSim not only aims at network measurements and UAV control, but also provides a comprehensive framework for the evaluation of IoT applications on UAVs. The end-to-end data-path enables the network simulator to receive and analyze sensor data from the UAVs and measure the QoS granted by different network scenarios and technologies. We observe that FlyNetSim can also integrate real-time processing of data, thus enabling the exploration of edge-assisted IoT architectures.

\begin{figure}[!t]
\centering
\includegraphics[width=\columnwidth]{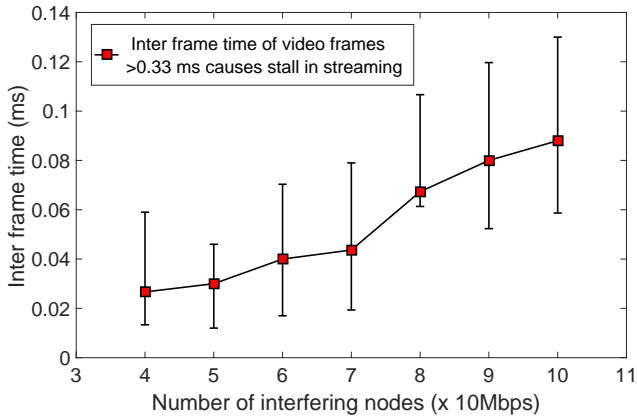}
\caption{Inter frame time of video frames at the decoder as a function of varying number of nodes contending the channel in the WiFi network.}
\label{fig:frame_delay}
\vspace{-1mm}
\end{figure}

\begin{figure}[!t]
\centering
\includegraphics[width=\columnwidth]{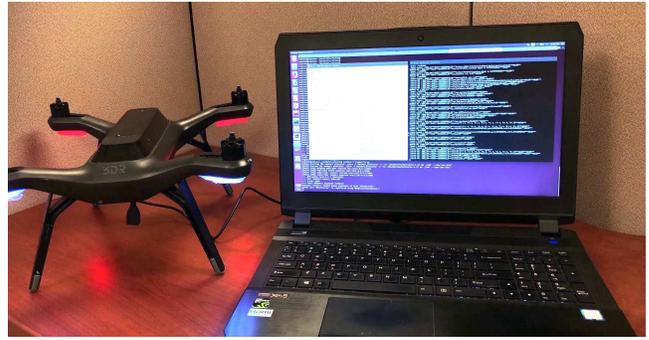}
\caption{Emulation mode where FlyNetSim is run on-board a 3DR Solo UAV.}
\label{fig:emu_uav}
\vspace{-1mm}
\end{figure}

We implement and test a scenario where FlyNetSim is used to support streaming of real-world encoded videos over the network simulator, where packet loss creates artifacts and impair the performance of processing algorithms. To the best of our knowledge, all available integrated UAV simulators do not support end-to-end encoded data transmission, especially over an articulated network infrastructure and a wide range of scenarios. In the experiments we used a H.264 AVC~\cite{wiegand2003overview} encoded video of $624$ frames with a resolution of $480$ pixels and frame rate of $30$ frames per second. We used the MPEG Transport Stream (TS) container for the transmission of the video. We set the network so that in the absence of contending nodes packet loss has very low probability and the video is delivered with perfect quality. Fig.~\ref{fig:psnr_vid} illustrates the degradation of the Peak Signal-to-Noise Ratio (PSNR) at additional nodes contending for the channel resource are introduced in the network. With $10$ additional nodes, the video stream incurs a PSNR degradation of more than $13$~dBs. Fig.~\ref{fig:frame_delay} shows the inter-frame time defined as the time lapse between consecutive frames. As the video in the experiment has a frame rate of $30$ frames per second, the expected average inter-frame time is $0.033$~ms. As the number of nodes contending the channel resource increases, the inter-frame time increases compared to the expected value.

%%%%%%%%%%%%%%%%%%%%%%%%%%%%%%
%%%%%%  Emulation Mode  %%%%%%
%%%%%%%%%%%%%%%%%%%%%%%%%%%%%%
\section{Emulation Mode}
\label{sec:emu}

\begin{figure}
\centering
\includegraphics[width=\columnwidth, height=0.6\columnwidth]{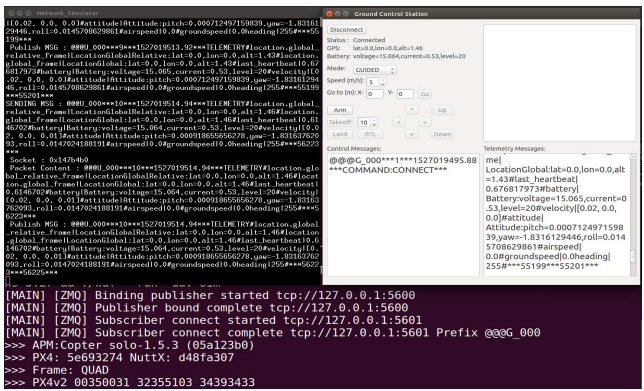}
\caption{Emulation telemetry output from 3DR Solo UAV}
\label{fig:emu_out}
\vspace{-1mm}
\end{figure}

FlyNetSim supports an ``emulation mode'', where a real UAV can communicate with external or simulated resources through the network simulator. This mode still uses all the components of the simulator, including the GCS and GUI panel, the middleware and NetSim, but the simulated UAV vehicle is replaced by a real-world UAV. We demonstrate the emulation mode by connecting a UAV 3DR Solo via USB serial to the laptop running the FlyNetSim as shown in Fig.~\ref{fig:emu_uav}. A similar configuration would enable on-board simulations during flight, for instance using a compact platform such as a Raspberry PI instead of the laptop.
Fig.~\ref{fig:emu_out} shows the GUI panel as telemetry regarding battery status, GPS location, etc., is received through the simulated network. The network simulator panel shows the log of the packet flow across the simulator from the UAV to the GCS. Note that the experiment was conducted indoor, where the GPS signal is not available, and thus the latitude, longitude values are shown as zero.

%%%%%%%%%%%%%%%%%%%%%%%%%%%
%%%%%%  Conclusions  %%%%%%
%%%%%%%%%%%%%%%%%%%%%%%%%%%
\section{Conclusions}
\label{sec:conclusions}
The primary contribution of this paper is a fully open source, integrated UAV-network simulator capable to support a wide range of network scenarios and use-cases while accurately modeling UAV operations. The architecture uses a lightweight custom built middleware to effectively simulate the UAV network with closed loop, synchronized, end-to-end data-paths, considerably reducing system resources usage compared to other available integrated simulators. Additionally, the support for emulation mode enables experimental research using real-world UAVs and sensor devices over simulated complex environments such as the urban IoT. We demonstrated the capabilities of the proposed simulator through a series of case study scenarios, including multi-technology networking, intra-swarm D2D communications and IoT data streaming.

\begin{acks}
%   The authors would also like to thank the anonymous referees for
%   their valuable comments and helpful suggestions. The work is
%   supported by the \grantsponsor{GS501100001809}{National Natural
%     Science Foundation of
%     China}{http://dx.doi.org/10.13039/501100001809} under Grant
%   No.:~\grantnum{GS501100001809}{61273304}
The work is partially supported by the \grantsponsor{IIS-1724331}{National
 Science Foundation}{http://dx.doi.org/10.13039/501100001809} under Grant
No.:~\grantnum{}{IIS-1724331}.
\end{acks}

\bibliographystyle{ACM-Reference-Format}
\balance
\bibliography{bibliography}

\end{document}